\documentclass[12pt]{iopart}
\usepackage{iopams}
\usepackage{graphicx}
\usepackage{verbatim}
\usepackage{bm}
\usepackage{bbm}
\usefont{T1}{tnr}{m}{sl}
\begin{document}
\title[Phase stability of cation-doped LiMnO$_{2}$]{Phase stability of cation-doped LiMnO$_{2}$ within the GGA+U
approximation}

\author{N N Shukla$^1$\footnote{Current Address: Electronic Research Administration, National Institutes of Health (NIH), 6705 Rockledge Dr., Bethesda, Maryland 20892, USA}, S. Shukla$^1$, R. Prasad$^1$ and R. Benedek$^2$}
\address{$^1$ Department of Physics, Indian Institute of Technology,
Kanpur-208016, India}
\ead{rprasad@iitk.ac.in}
\address{$^2$ Chemical Sciences and Engineering Division, Argonne National Laboratory,
 Argonne, Illinois 60439, USA}

\begin{abstract}
First principles density functional theory calculations within the GGA+U approximation were performed for LiMn$_{1-x}$M$_x$O$_{2}$, a candidate cathode material for lithium-ion batteries, with ($x=0.25$, M=Ni, Fe, Co, Mg), to investigate the effect of doping on the destabilization of the monoclinic structure relative to the layered rhombohedral structure. A primary motivation of this work was to determine to what extent the predictions of the electronically more realistic GGA+U treatment would differ from those obtained within the GGA. Several significant qualitative changes are found. For the pristine system in the rhombohedral structure, Mn ions show a high-spin state within GGA+U, rather than the low spin (metallic) state found in GGA. The doped rhombohedral structure is unstable within GGA+U, rather than metastable, as in GGA. In the monoclinic structure, the dopant oxidation states are the same (trivalent Fe, divalent Co, Ni) in GGA+U and GGA. Co and Ni ions show a higher spin state in GGA+U than in GGA. The divalent dopants destabilized the monoclinic structure to a greater extent than trivalent dopants within the GGA+U, as expected from previous GGA calculations. Overall, our results suggest that the relatively close agreement sometimes found between properties calculated within the GGA and GGA+U may be misleading, because the underlying electronic behaviors may be profoundly different.
\end{abstract}



\section{Introduction}
Among the most promising cathode materials for secondary lithium-ion 
batteries \cite{tar,arm,pet} are lithiated transition metal oxides. To model the 
electronic and structural properties of such materials realistically  from first principles has been technically challenging, in view of the shortcomings of self-consistent-field treatments of the strongly correlated electronic states in many transition-metal oxides, at the level of the local-density (or local-spin-density) approximation. It has long been known, for example, that local-density-functional theory methods typically predict gapless (metallic) electronic spectra for insulating oxides \cite{tera}. Despite the unrealistic electronic spectra predicted by local density 
functional theory, integral properties that depend on total energy, which involve 
summations over all occupied electronic states, such as crystal structure parameters, are 
nonetheless often predicted with reasonable accuracy within local spin density functional theory. Even the structural properties, however, may not be properly modeled, if the local-spin-density approximation treatment does not capture the correct underlying electronic structure, e.g., the transition metal oxidation state, at least qualitatively.   

In previous work \cite{rp1}, the GGA was applied to calculate some electronic and structural properties of layered LiMnO$_2$ with substitutional dopants, an interesting class of candidate lithium-ion battery materials. Of particular interest is the ability of divalent dopants to promote 
stabilization of the layered rhombohedral structure by depopulating the majority-spin Mn $e_g$ states and thereby destabilize the Jahn-Teller-distorted monoclinic structure to a greater extent than trivalent dopants. We anticipate that a trend from trivalency to divalency occurs from early to late transition element solutes. This simple picture was borne out reasonably well (cf. Fig. (9) in our previous paper \cite{rp1}) by GGA calculations. The stabilization of the rhombohedral relative to the monoclinic structure showed an increasing trend, although with some scatter, for dopants across the first transition series, with the early transition elements adopting a more trivalent character, and the late transition elements a more divalent character. 

Although this picture, in which the solute oxidation state is the property of primary importance, is appealing in its simplicity, the inability of GGA, in general, to predict physically accurate electronic spectra of correlated transition metal oxides raises doubts about its accuracy in predictions of spectrally sensitive properties such as oxidation states. It therefore seems desirable to see how well the results of the GGA treatment hold up in higher level calculations.

Among higher-level mean-field treatments, the LSDA-SIC \cite{petit}, GW \cite{kotani} and LSDA+U (or GGA+U) \cite{aniso1,ldau} methods come into consideration. We have employed in this work the GGA+U method, which is a relatively simple extension of the GGA method employed in our previous work \cite{rp1}, with essentially the same computational burden. LSDA+U 
improves the electron-energy-loss spectra and parameters characterizing the structural 
stability of NiO \cite{duda,aniso1}, compared to LSDA. Improved results of LSDA+U predictions for some other oxides \cite{vladi,ceder,roll,zhou} further demonstrate its merit. 

In this work, we have applied the GGA+U to treat several of the dopants of layered LiMnO$_2$ considered in our earlier work \cite{rp1}, to determine whether the dopants show dramatically different behavior within this treatment. We find electronic structures that are considerably different for the undoped rhombohedral structure, and for the Co and Ni doped materials with monoclinic structure.  In addition to our calculations for the monoclinic and rhombohedral structures, results are presented for the (undoped) orthorhombic structure, not treated in our previous work. 

The method of computation is briefly reviewed in Sec. \textbf{II}, section \textbf{III} presents results, and conclusions are given in Sec. \textbf{IV}.

\section{Computational details}
\subsection{Method}
We employ in this work the projector augmented wave (PAW) method, as implemented 
in the Vienna ab initio simulation package (VASP) \cite{wien}, with the GGA 
(generalized gradient correction) parametrization given by Perdew et al \cite{john}. In the 
GGA+U extension of density functional theory\cite{aniso1,ldau}, an effective on-site 
Coulomb potential for the $d$-states is added to the GGA \cite{duda}. For comparison, calculations were also done at the GGA level, using the PAW implementation. In  previous work \cite{rp1}, GGA calculations were performed using ultrasoft pseudopotentials. 

The integration over the Brillouin zone is performed using special k-points generated 
with Monkhorst-Pack\cite{pack} indices (4,4,4). The optimized geometry
is finally calculated for higher numbers of k-points using indices (8,8,8) and a high 
plane-wave energy cutoff of 550 eV to ensure good convergence of 
stress tensor during cell-parameter relaxations. For each system both the volume and 
internal coordinate relaxations are performed during the self consistent energy 
minimization. An improved tetrahedron method has been used for the Brillouin-zone 
integration \cite{blochl}. We select a Hubbard parameter $U$ of 6 eV for Mn and for the dopants Fe, 
and Co and 7 eV for Ni,  guided by Zhou ${\it et} {\it al.}$ \cite{zho}. The exchange energy $J$ is fixed at 1 eV \cite{jvalue}.

\subsection{Structure}
Calculations are performed for the orthorhombic, monoclinic, and rhombohedral 
structures of LiMnO$_2$, whose space groups are Pmmn, C2/m and R$\bar{3}$m, 
respectively. (GGA calculations only for the monoclinic and rhombohedral structures were presented in our earlier work \cite{rp1}.) The experimental lattice constants for the orthorhombic \cite{ortho1} and 
monoclinic\cite{arm} structures are taken as initial values, prior to optimization. 
For the hypothetical rhombohedral structure of pure LiMnO$_2$, we take as starting 
parameters the lattice constants calculated by Mishra and Ceder \cite{mishra}, and a ferromagnetic spin configuration. 
For the orthorhombic and monoclinic structures, spin-polarized calculations have been performed for both ferromagnetic (FM) and 
antiferromagnetic (AFM) arrangements of the Mn spin. The corresponding magnetic 
structural unit cells are constructed as described by Prasad ${\it et} {\it al.}$ \cite{rp1}. 
In the monoclinic structures, the AF3 ordering proposed by 
Singh\cite{singh} has been used for the AFM case. Supercells with four formula units for 
the monoclinic, rhombohedral structures and eight formula units for orthorhombic 
structure were employed. 

In calculations for doped materials, we focus on the monoclinic rather than the orthorhombic structure, although experimental investigations of the doped orthorhombic system have also been done \cite{myung06}. The reason the monoclinic system was investigated in this work is its similarity to the rhombohedral structure, of which it is a distorted version. This similarity suggests a way to analyze the relative tendency of different dopants to destabilize the monoclinic distortion.  With sufficient doping, the orthorhombic structure would also likely transform to rhombohedral \cite{ammun00}. 

\section{Results and Discussion}
\subsection{Pristine System}
Equilibrium lattice parameters, calculated within the GGA+U approximation, for undoped orthorhombic, 
monoclinic and rhombohedral LiMnO$_2$, are listed in Table I. Results obtained in 
previous work are also given, for comparison. The largest differences occur for the
rhombohedral structure, compared to previous GGA calculations\cite{mishra}.
The overall ground-state is found to be AF orthorhombic, in agreement with earlier 
LSDA calculations\cite{mishra}. Longer Mn-O bond lengths are obtained in all cases, compared 
with earlier LSDA or GGA calculations\cite{mishra}, which is a general feature of 
GGA+U. The underestimation of lattice constants (and bond lengths) within the LSDA is 
well known, particularly for metals with high compressibility\cite{john}. The 
dependence of structural parameters on $U$ can be understood from the weakening of covalent bonding with increasing correlation, which shifts the distribution of electronic charge density from the interstitial to the atomic regions, and results in weak orbital hybridization in the presence of strong coulomb correlation\cite{phil}.

The total and partial Mn d-electron density of states (DOS) in the monoclinic (AFM) 
structure is shown in Fig. 1. The curves above and below the energy axis represent the 
majority and minority carrier states, and the vertical dash-dotted line at zero on the 
abscissa indicates the Fermi energy. The top-most panel, which shows the total DOS, 
agrees with our earlier work\cite{nit}. Our calculated bandgap, about 1.5 eV, is slightly 
higher than obtained in a previous\cite{gala} LSDA+U calculation for this material.

The lower panels show site-projected densities of states. The Fermi level appears between 
the occupied and unoccupied Jahn-Teller split $e_g$ states of Mn. The minority states on 
the Mn site are almost entirely unoccupied. The last two panels show the O-p
DOS, which appears at the top of the valence band\cite{nit}.

It was noted in earlier work \cite{rp1,zfw} that Mn ions adopt a low spin state in rhombohedral LiMnO$_2$ within the GGA. In contrast, GGA+U shows a high spin state for trivalent Mn in rhombohedral LiMnO$_2$, as would be expected from its behavior in other materials. Rhombohedral LiMnO$_2$, however, is structurally unstable within the GGA+U; only by distorting the structure to a lower (e.g., monoclinic) symmetry can it be stabilized. By performing calculations as a function of $U-J$, we find that the low spin state, and stabilization of the rhombohedral structure, occurs for $U-J$ less than about 2.3 eV. Since undoped rhombohedral LiMnO$_2$ is unstable within the GGA+U, the analysis performed in our earlier work \cite{rp1}, in which low-spin rhombohedral LiMnO$_2$ was found stable within the GGA,  is not applicable. We therefore confine our doped LiMnO$_2$  GGA+U calculations in this work to the monoclinic structure. 

\subsection{Doped systems}
Our previous calculations within the GGA suggested that the relative stability of the rhombohedral structure may be promoted by the introduction of 
dopants with trivalent, and particularly divalent oxidation states, which suppress the cooperative Jahn-
Teller effect\cite{prasad,rp1}. In this work, we investigated the substituents Ni, Fe, Co and 
Mg on the Mn sublattice in AFM LiMnO$_2$, in the presence of coulomb repulsion parameter $U$.

Fig. 2 shows the Mn and Ni d-DOS for LiNi$_{0.25}$Mn$_{0.75}$O$_2$.
The top-most panel (a) shows the majority- and minority-spin DOS for Ni, and panels
(b), (c) and (d) show the majority and minority DOS for the three Mn ions in the four 
formula-unit supercell used in our calculations. The empty majority $e_g$ band in panel 
(b), in contrast with the Mn $d$-DOS for pure LiMnO$_2$ (Fig. 1), reveals the oxidation of the Mn to the 4+ state in the presence of the Ni ion. Confirming that the Mn ion represented in panel (b) is tetravalent is the surrounding oxygen octahedron, which shows essentially no Jahn-Teller distortion, unlike the other Mn ions in the cell.  Panels (c) and
(d) show the DOS of Mn$^{3+}$ ions, similar to Mn ions in the
undoped system, shown in middle panel of Fig. 1. The majority Ni $d$-DOS in panel (a)
shows fully occupied $t_{2g}$ and e$_g$ bands while minority DOS shows fully
occupied $t_{2g}$ and empty $e_g$ bands, consistent with a divalent oxidation state. 
A similar analysis of site-projected DOS, as well as oxygen octahedra, for the doped systems 
LiCo$_{0.25}$Mn$_{0.75}$O$_2$,
LiFe$_{0.25}$Mn$_{0.75}$O$_2$, and LiMg$_{0.25}$Mn$_{0.75}$O$_2$ 
reveals that Fe adopts the 3+ oxidation state whereas Co and Mg are 2+.

In the case of Co, divalent or trivalent oxidation states can be obtained, depending on the initial atomic arrangement, prior to relaxation of the internal coordinates. If the Jahn-Teller distortion of the oxygen octahedron surrounding one of the Mn ions is suppressed in the initial configuration, the relaxed calculation yields a tetravalent state for than Mn, and a divalent state of Co; otherwise, all transition ions are trivalent. We find that the state in which Co is divalent is lowest in energy, and the state in which Co is trivalent is therefore metastable. 

\subsection{Destabilization of monoclinic  structure by dopants} 
Since the rhombohedral structure of LiMn$O_2$ is unstable, within the GGA+U approximation, we perform a different analysis from that in our previous work \cite{rp1} to assess the destabilization of the monoclinic structure with doping. We note that the layered rhombohedral structure is a special case of the monoclinic structure that occurs when 
\begin{equation}
a/b=\surd 3,
c \cos \beta= -a/3,
\end{equation}
where $a$, $b$, and $c$ are the lattice constants and $\beta$ is the monoclinic angle \cite{rp1}.  In the pristine system,
\begin{equation} 
h(x=0) \equiv a/(\surd 3 b) = 1.12,
\end{equation}
When dopants are introduced, the relaxed lattice constants $a(x)$, $b(x)$, and $c(x)$ vary. Consequently $h$ decreases, and, at a critical concentration that corresponds to the phase boundary between monoclinic and rhombohedral structures, decreases, perhaps discontinuously, to unity. (A two-phase coexistence region is also possible, in principle. For the sake of discussion, we also ignore solubility limits on $x$).

In Fig. 3, we plot the quantity
\begin{equation}
H(x) \equiv [h(x)-h(0)]/[1-h(0)],
\end{equation}
evaluated with lattice constants calculated within the GGA+U approximation. The function $H(x)$ is, by construction, zero at $x=0$ and unity at the critical concentration, $x_{cr}$.  The more effective a dopant is in destabilizing the monoclinic structure relative to the rhombohedral structure, the larger the value of $H(x)$. (We note that this analysis does not directly yield a prediction of the critical dopant concentration.) As shown in the Figure, the divalent dopants Mg, Ni and Co are more effective than the trivalent dopant, Fe. This is consistent with trends observed in previous work. 

We have also calculated amplitudes of Jahn-Teller modes  for Mn- and dopant centered octahedra in the monoclinic structure. Our values are close to the corresponding values of our previous work\cite{rp1} indicating that the dopants reduce the Jahn-Teller distortion in the GGA+U also.

\section{Summary and Conclusions}
Little difference was found between GGA+U and GGA calculations of some structural properties of undoped and doped LiMnO$_2$.
Lattice parameters calculated for orthorhombic and monoclinic structures of pure 
LiMnO$_2$ within the GGA+U framework are in reasonable agreement with earlier 
work based on the GGA. Consistent with previous GGA-based 
calculations\cite{mishra}, as well as experiment \cite{ortho2}, the orthorhombic structure 
in the AF configuration is found to be the overall ground state in the GGA+U approximation. 

Significant qualitative differences between electronic structures calculated in GGA+U and GGA were found in some cases, however, which indicate that the agreement cited above may be misleading. For rhombohedral LiMnO$_2$, GGA+U gives a high-spin solution, in contrast to the low spin result found in GGA \cite{rp1,zfw}. Furthermore, the rhombohedral structure in GGA+U is not metastable, as in GGA, but unstable. Although it cannot be verified experimentally, because rhombohedral LiMnO$_2$ is a hypothetical structure, the electronic structure calculated within the GGA for (low-)doped or undoped  rhombohedral LiMnO$_2$ are assumed to be unphysical. To treat doping in the present work, we have therefore focused on the monoclinic structure, and the changes induced by doping. For Ni and Co dopants, divalent oxidation states are found both in GGA and GGA+U, but GGA+U gave a significantly higher state of polarization.  In the case of Fe dopants, a high spin state is predicted both by GGA and GGA+U.

The greater destabilization of the monoclinic structure induced by divalent than by trivalent dopants previously found within the GGA, was found to hold also within GGA+U.

\ack
This work was supported at Argonne National Laboratory by the office of FreedomCAR 
and Vehicle Technologies(Batteries for Advanced Transportation Technologies(BATT) Program),
U.S. Department of Energy. Argonne National Laboratory, a U.S Department of Energy Office of Science Laboratory, is operated by UChicago Argonne,
LLC, under contract no. DE-AC02-06CH11357. Grants of computer time at the 
National Energy Research Supercomputer Center, Lawrence Berkeley Laboratory, 
are gratefully acknowledged.

\newpage
TABLE I. Lattice parameters of monoclinic (MLA),
orthorhombic (ORTHO) and rhombohedral (RLA) structures of LiMnO$_2$ for AF and FM 
configurations, based on GGA+U approximation. The values given in parentheses are results of previous
LSDA(GGA) calculations
 \cite{mishra} and the available experimental work \cite{arm,ortho1} respectively.\\
\\
\begin{tabular}{lcrrrrr}\hline
\multicolumn{1}{c}{Structure} &
\multicolumn{1}{c}{a (\AA)} &
\multicolumn{1}{c}{b (\AA)} &
\multicolumn{1}{c}{c (\AA)} &
\multicolumn{1}{c}{$\beta$} &
\multicolumn{1}{c}{internal parameter } \\ \hline

MLA(AF)&5.49&2.86&5.44&112.87$^\circ$&x=0.271, z=0.769\\
               &(5.54)&(2.77)&(5.47)&(116.0$^\circ$)&(x=0.271, z=0.762)(Ref. 23)\\
               &(5.44)&(2.81)&(5.39)&(116.0$^\circ$)&(x=0.272, z=0.771)(Ref. 2)\\
~~~~~~~~     (FM)&5.48&2.87&5.43&112.85$^\circ$&x=0.270, z=0.769\\
               &(5.54)&(2.82)&(5.44)&(116.0$^\circ$)&(x=0.270, z=0.763)(Ref. 23)\\
ORTHO(AF)&2.87&4.60&5.81&90.0$^\circ$&z$_{Li}$=0.117, z$_{Mn}$=0.636 \\
& & & & &z$_{0}$=0.139, 0.600  \\
               &(2.79)&(4.69)&(5.64)&(90.0$^\circ$)&(z$_{Li}$=0.104, 
z$_{Mn}$=0.636)(Ref. 20)\\
               & & & & &(z$_{0}$=0.132, 0.606)\\
               &(2.81)&(4.57)&(5.76)&(90.0$^\circ$)&(z$_{Li}$=0.126, 
z$_{Mn}$=0.635)(Ref. 22)\\
~~~~~~~~     (FM)&2.87&4.61&5.81&90.0$^\circ$&z$_{Li}$=0.117, 
z$_{Mn}$=0.637 \\
             & & & & &z$_{0}$=0.138, 0.600\\
               &(2.80)&(4.82)&(5.60)&(90.0$^\circ$)&(z$_{Li}$=0.116, 
z$_{Mn}$=0.638)(Ref. 23)\\
               & & & & &(z$_{0}$=0.128, 0.605)\\
               & & & & &(z$_{0}$=0.144, 0.602)\\
RLA(FM)&3.02&3.02&14.58&90.0$^\circ$&x=0.257\\
                 &(2.82)&(2.82)&(14.27)&(90.0)$^\circ$&(x=0.255)(Ref. 23)\\ \hline
\end{tabular}\\
\\
\\

\newpage
$\bf{FIGURE}$ $\bf{CAPTIONS}$

Fig. 1. The total DOS and the partial Mn-d and O-2p DOS of the monoclinic LiMnO$_2$ 
in antiferromagnetic state calculated at optimized lattice parameters. Upper most panel 
shows total DOS. The middle and lower panels show partial Mn-d and O-2p DOS 
respectively.

Fig. 2. Partial d-DOS of AFM monoclinic LiNi$_{0.25}$Mn$_{0.75}$O$_2$. Panel (a) 
shows d-DOS of Ni$^{2+}$. Panel (b), (c) and (d) show d-DOS for Mn ions that are 
neighbors of Ni dopant. One Mn adopts 4+ oxidation state, while other two remain in the 
3+ oxidation state.

Fig. 3. The function $H(x)$ at $x=0.25$ for dopants Mg, Ni, Co and Fe, calculated within GGA+U approximation.

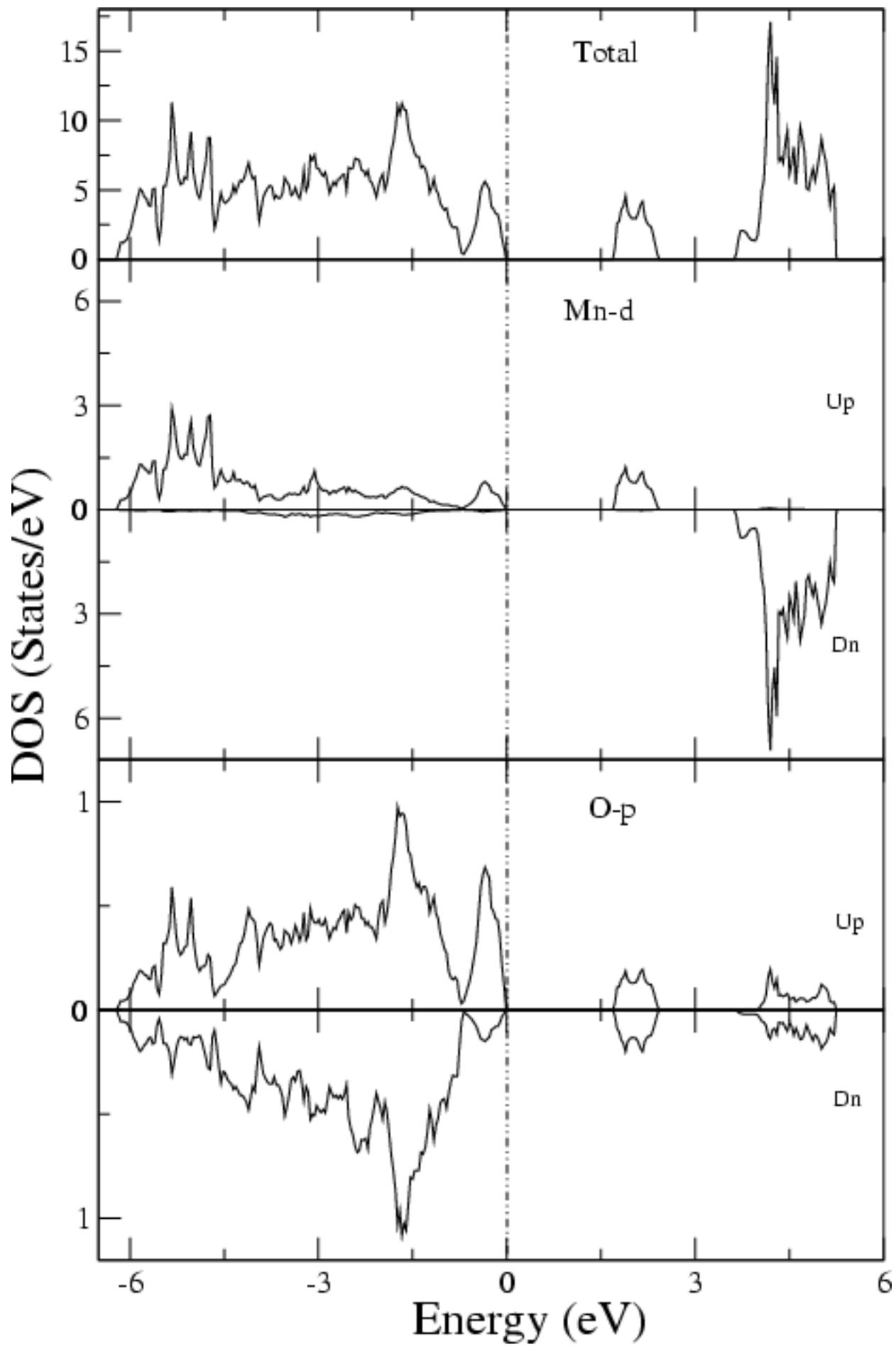
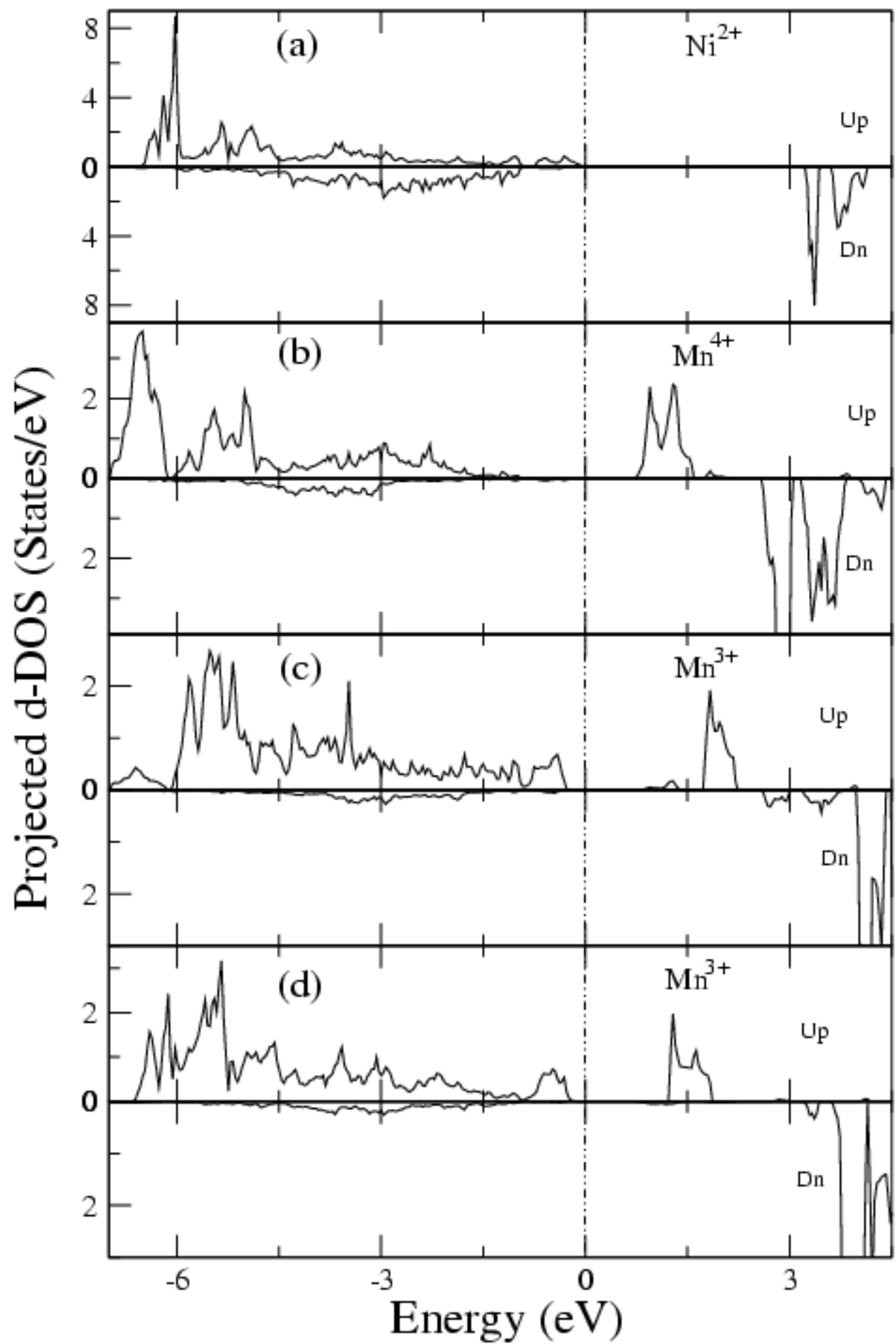
\end{document}